\documentclass[aps,prb,reprint,groupedaddress]{revtex4-1}
\usepackage{graphicx}
\usepackage{color,amsmath}

\usepackage[normalem]{ulem}

\begin{document}


\title{Edge Transport in the Trivial Phase of InAs/GaSb}



\author{Fabrizio Nichele}
\affiliation{Center for Quantum Devices and Station Q Copenhagen, Niels Bohr Institute, University of Copenhagen, Universitetsparken 5, 2100 Copenhagen, Denmark}
\email[email: ]{fnichele@nbi.ku.dk}

\author{Henri J. Suominen}
\affiliation{Center for Quantum Devices and Station Q Copenhagen, Niels Bohr Institute, University of Copenhagen, Universitetsparken 5, 2100 Copenhagen, Denmark}

\author{Morten Kjaergaard}
\affiliation{Center for Quantum Devices and Station Q Copenhagen, Niels Bohr Institute, University of Copenhagen, Universitetsparken 5, 2100 Copenhagen, Denmark}

\author{Charles M. Marcus}
\affiliation{Center for Quantum Devices and Station Q Copenhagen, Niels Bohr Institute, University of Copenhagen, Universitetsparken 5, 2100 Copenhagen, Denmark}

\author{Ebrahim Sajadi}
\affiliation{Quantum Matter Institute, University of British Columbia, Vancouver, BC, V6T1Z4, Canada}
\affiliation{Department of Physics and Astronomy, University of British Columbia, Vancouver, BC, V6T1Z1, Canada}

\author{Joshua A. Folk}
\affiliation{Quantum Matter Institute, University of British Columbia, Vancouver, BC, V6T1Z4, Canada}
\affiliation{Department of Physics and Astronomy, University of British Columbia, Vancouver, BC, V6T1Z1, Canada}

\author{Fanming Qu}
\affiliation{QuTech and Kavli Institute of Nanoscience, Delft University of Technology, 2600 GA Delft, The Netherlands}

\author{Arjan J.A. Beukman}
\affiliation{QuTech and Kavli Institute of Nanoscience, Delft University of Technology, 2600 GA Delft, The Netherlands}

\author{Folkert K. de Vries}
\affiliation{QuTech and Kavli Institute of Nanoscience, Delft University of Technology, 2600 GA Delft, The Netherlands}

\author{Jasper van Veen}
\affiliation{QuTech and Kavli Institute of Nanoscience, Delft University of Technology, 2600 GA Delft, The Netherlands}

\author{Stevan Nadj-Perge}
\affiliation{QuTech and Kavli Institute of Nanoscience, Delft University of Technology, 2600 GA Delft, The Netherlands}

\author{Leo P. Kouwenhoven}
\affiliation{QuTech and Kavli Institute of Nanoscience, Delft University of Technology, 2600 GA Delft, The Netherlands}

\author{Binh-Minh Nguyen}
\affiliation{HRL Laboratories, 3011 Malibu Canyon Road, Malibu, California 90265, USA}

\author{Andrey A. Kiselev}
\affiliation{HRL Laboratories, 3011 Malibu Canyon Road, Malibu, California 90265, USA}

\author{Wei Yi}
\affiliation{HRL Laboratories, 3011 Malibu Canyon Road, Malibu, California 90265, USA}

\author{Marko Sokolich}
\affiliation{HRL Laboratories, 3011 Malibu Canyon Road, Malibu, California 90265, USA}

\author{Michael J. Manfra}
\affiliation{Department of Physics and Astronomy and Station Q Purdue, Purdue University, West Lafayette, Indiana 47907 USA}
\affiliation{School of Materials Engineering, Purdue University, West Lafayette, Indiana 47907 USA}
\affiliation{School of Electrical and Computer Engineering, Purdue University, West Lafayette, Indiana 47907 USA}
\affiliation{Birck Nanotechnology Center, Purdue University, West Lafayette, Indiana 47907 USA}

\author{Eric M. Spanton}
\affiliation{Stanford Institute for Materials and Energy Sciences, SLAC National Accelerator Laboratory, Menlo Park, California 94025, USA}
\affiliation{Department of Physics, Stanford University, Stanford, California 94305, USA}

\author{Kathryn A. Moler}
\affiliation{Stanford Institute for Materials and Energy Sciences, SLAC National Accelerator Laboratory, Menlo Park, California 94025, USA}
\affiliation{Department of Physics, Stanford University, Stanford, California 94305, USA}
\affiliation{Department of Applied Physics, Stanford University, Stanford, California 94305, USA}


\date{\today}

\begin{abstract}
We present transport and scanning SQUID measurements on InAs/GaSb double quantum wells, a system predicted to be a two-dimensional topological insulator. Top and back gates allow independent control of density and band offset, allowing tuning from the trivial to the topological regime. In the trivial regime, bulk conductivity is quenched but transport persists along the edges, superficially resembling the predicted helical edge-channels in the topological regime. We characterize edge conduction in the trivial regime in a wide variety of sample geometries and measurement configurations, as a function of temperature, magnetic field, and edge length. Despite similarities to studies claiming measurements of helical edge channels, our characterization points to a non-topological origin for these observations.
\end{abstract}


\maketitle

\section{Introduction}
Quantum spin Hall (QSH) insulators are topologically non-trivial two-dimensional materials characterized by an insulating bulk and helical modes at the sample edges  \cite{Bernevig2006}. Among two-dimensional systems predicted to exhibit a QSH insulating phase, the InAs/GaSb double quantum well (QW) system is especially promising for device applications \cite{Liu2008,Knez2011,Suzuki2013,Nichele2014,Knez2014,Spanton2014,Du2015,Qu2015}. Compared to inverted HgTe/HgCdTe QWs, where the QSH effect was first reported \cite{Konig2007,Roth2009}, the InAs/GaSb system offers high mobility and ease of fabrication characteristic of III-V heterostructures, and an electrically tunable band structure. In particular, by the combined action of top and back gates, the Fermi level position and the overlap between the InAs conduction band and the GaSb valence band can be independently controlled \cite{Liu2008, Qu2015}. In that way, the system can be tuned from a trivial insulating phase, similar to a conventional semiconductor, to the inverted regime, with a hybridization gap between valence and conduction bands marking the QSH phase. Inverted (topological) and non-inverted (trivial) band alignments are schematically represented in Fig.~\ref{fig:delft1}(a) left and right panels. 

Early experimental evidence of edge-channel conduction in InAs/GaSb QWs was reported in micron-sized samples in Ref.~\onlinecite{Knez2011}. Subsequent refinements involved adding Si impurities at the interface between quantum wells \cite{Spanton2014,Knez2014,Du2015} or using Ga sources of reduced purity \cite{Charpentier2013,Mueller2015} to quench residual bulk conduction. These reports convincingly establish that conducting edges are robustly observed in the InAs/GaSb system.

Missing from previous work were critical tests that establish that the observed conducting edges are indeed the helical modes predicted to exist at the boundary of a 2D topological insulator. For instance, helical edges are expected to have a length-independent quantized conductance for sample lengths shorter than a characteristic spin scattering length. While observed edge-channel conductances were close to expected values \cite{Knez2014, Du2015, Mueller2015}, the crossover from a length-dependent conductance for long samples to length independent quantized conductance for short samples was not demonstrated. Moreover, because the crossover from trivial to topological regimes was not mapped out, the observed edge-channel conduction should be taken as circumstantial rather than direct evidence for helical edge states and hence the topological phase. In particular, Fermi level pinning at the surface or other effects that can give rise to edge conduction were not subject to experimental test. 

We previously showed how the electronic phase of our samples can be tuned \textit{in situ} from the trivial to the inverted regime, and how the bulk phases can be distinguished \cite{Qu2015}. In this paper we extend the study to the edges of our samples. By combining transport methods with spatially resolved scanning superconducting quantum interference device (SQUID) measurements, we map the edge channel behavior in the trivial and inverted electronic phase of InAs/GaSb. The central conclusion we reach from the collection of measurements presented here is that when the sample is tuned into the \textit{trivial} regime, conductance is suppressed through the bulk but remains along the sample edges. We emphasize that  edge conduction is observed in the trivial regime, where helical states are not expected. At a superficial level, the edge conduction characteristics we observed are similar to those reported previously as evidence for the QSH state in InAs/GaSb.

Plateaus in resistance at apparently quantized values [Sec.~\ref{microsamples}] are observed in an H bar geometry that was designed to resemble devices described in existing literature \cite{Roth2009,Du2015}. This result alone is not sufficient to prove the presence of helical edge channels. We therefore complement it with additional samples aimed at identifying the topological phase in the bulk (either trivial of inverted), the residual bulk conductance and the nature of the edge channels (helical, ballistic or diffusive).
Residual bulk transport in the inverted regime of our samples makes the detection of any edge-channel conduction difficult via conventional transport measurement. However, scanning probe techniques demonstrate the existence of edge channels also in the inverted regime, with similarities to those measured in the trivial regime.
We find that in the trivial regime the edge resistance scales linearly with edge length even in the limit of very short edges, contrary to the expectation for quantized helical edges. Furthermore, the edge channel resistance per unit length is very close to earlier reports of helical edge channels \cite{Roth2009,Du2015}.
These observations imply a burden on future QSH experiments in InAs/GaSb to confirm not only the helical character of edges in the inverted regime, but also the absence of edge transport in the trivial regime that might otherwise conduct in parallel with helical modes.
		
The paper is organized as follows: First, details of sample fabrication and measurements are provided. Macroscopic transport measurements in Hall bar and Corbino geometries map out trivial and inverted regimes of gate voltage. Taken together, these measurements show that conduction in the trivial regime is entirely along the sample edges, with an immeasurably small contribution from the bulk. The length dependence of the edge resistance is measured using mesoscopic two-terminal devices. The resistance falls well below the expected $h/2e^2$ for edge segments shorter than one micron. We proceed with an investigation of H bars and microscopic ($\mu$) Hall bars with dimensions very similar to those reported in earlier work \cite{Roth2009,Du2015}. Here we note the remarkable coincidence that typical edge resistivity in these samples gives resistances near those expected from quantization for the same sample geometries and sizes reported in the literature, despite the fact that our measurements are manifestly performed in the trivial regime. We then demonstrate edge conduction through the entire phase diagram with a scanning probe technique. Enhanced conduction at the sample edge is also seen in the inverted regime, but there it competes with a significant bulk contribution. In Sec.~\ref{sec:discussion} we mention different scenarios to account for the origin of the edge channels and propose experimental ways to suppress their contribution. 

From the key observations of this paper, namely:
\begin{itemize}
\item A pronounced edge channel conduction exists in InAs/GaSb in the trivial regime.
\item The two-terminal resistance of an edge channel linearly scales with length, taking values smaller than $h/e^2$ for short edges.
\item The newly discovered edge channels have an insulating temperature dependence and a weak dependence on an in-plane magnetic field.
\item The typical edge channel resistivity is so that resistance values close to $h/e^2$ can be obtained for sample sizes and geometries similar to those reported in previous work.
\end{itemize}
We conclude that previous and future experiments on quantum spin Hall materials must be tested against spurious sources of edge-channel conduction.

\section{Experimental Details \label{expdetails}}

Experiments were performed on three different wafers, labeled A, B, and C. The structures were grown by molecular beam epitaxy on a conductive GaSb substrate, which served as a global back gate \cite{Nguyen2015}. From the substrate to the surface, all three structures consisted of a GaSb/AlSb insulating buffer, a $5~\rm{nm}$ GaSb QW, an InAs QW ($10.5~\rm{nm}$ for wafers A and B, $12.5~\rm{nm}$ for wafer C), a $50~\rm{nm}$ AlSb insulating barrier and a $3~\rm{nm}$ GaSb capping layer. Transport experiments were performed on wafers A and B, although measurements reported here (Figs~\ref{fig:delft1}-\ref{fig:Hbar}) are from wafer A only. Analogous measurements on wafer B gave consistent results. Scanning SQUID measurements (Figs.~\ref{fig:SQUID} and~\ref{fig:SQUID2}) were performed on Wafer C, previously characterized by transport measurements in Ref.~\onlinecite{Qu2015}. Magnetotransport measurements reported here and elsewhere \cite{Qu2015} confirm that for wafers A, B, and C, the band structure is trivial (non-inverted) at $V_{\rm{BG}}=0$.

Material quality is reflected in a higher electron mobility than material used in previous reports \cite{Knez2011,Suzuki2013,Nichele2014,Knez2014,Spanton2014,Du2015}. The mobility versus density characteristic of wafer C was measured in Refs.~\onlinecite{Nguyen2015,Qu2015}, yielding mobility values in excess of $50~\rm{m^2V^{-1}s^{-1}}$ for an electron density of $10^{16}~\rm{m^{-2}}$. The mobility in wafer A and B follows a similar dependence on density as wafer C, with an overall decrease by about a factor of two.

We adopt very similar fabrication recipes as in previous edge channels studies in InAs/GaSb \cite{KnezThesis,Suzuki2013,Pal2015}. Devices were patterned by conventional optical and electron beam lithography and wet etching. Devices shown in Figs.~\ref{fig:delft1} and \ref{fig:delft2} were etched using a sequence of selective etchants \cite{Yang2002}, the other devices with a conventional III-V semiconductor etchant \cite{Nguyen2015}. The two recipes gave consistent results. Ohmic contacts were obtained by etching the samples down to the InAs QW and depositing Ti/Au electrodes. Top gates were defined by covering the samples with a thin ($\leq 80~\rm{nm}$) $\rm{Al_{2} O_{3}}$ or $\rm{HfO_2}$ insulating layer grown by atomic layer deposition and a patterned Ti/Au electrode. The one exception to this was the Corbino disk presented in Fig.~\ref{fig:delft2}, for which the insulator consisted of a $90~\rm{nm}$ sputtered layer of $\rm{Si_3N_4}$.

Special care was taken during the entire fabrication process not to accidentally create or enhance spurious edge conductance in the samples. In particular it is known that antimony compounds react with oxygen and optical developpers giving rise to amorphous conductive materials \cite{Gatzke1998, Chaghi2009}. We therefore always store the samples in nitrogen, never heat the samples above $180~\mathrm{^{\circ}C}$ and deposit the insulating oxides immediately after the wet etching, serving as a passivating layer. The exclusively use of electron beam lithography allowed us to avoid optical developers.

In many devices, the back gate leaked when more than $\pm 100{\rm mV}$ was applied, presumably due to damage during processing. These leaky devices were only operated at zero backgate voltage, where the resistance to the backgate was at least $10~\rm{G\Omega}$.
Except where specified, transport experiments were performed in dilution refrigerators at a temperature of less than $50~\rm{mK}$ with standard low frequency lock-in techniques. Additional details regarding wafer growth, sample fabrication, and basic electrical characterization are provided in Refs.~\onlinecite{Nguyen2015,Qu2015}.

\section{Transport in Macroscopic Samples \label{macro}}
\subsection{Magnetotransport Data\label{magnetotransport}}
\begin{figure}
\includegraphics[width=\columnwidth]{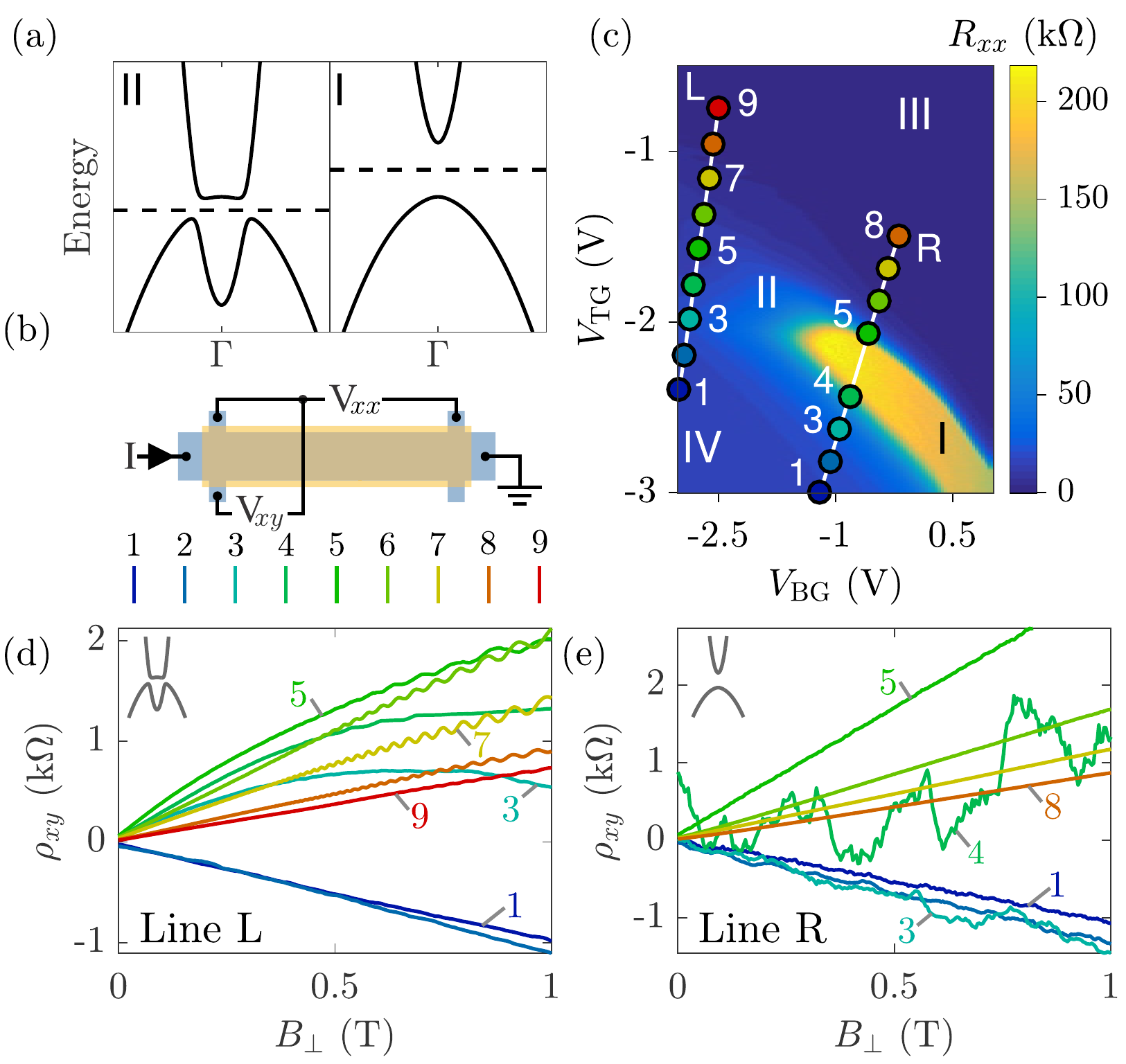}
\caption{(a) Schematic representation of the InAs/GaSb band structure for inverted (left) and trivial (right) regime. We interpret region II and region I in (c) as the situation when the Fermi energy (dashed line) lies in the gap in the inverted and trivial case, respectively. Through the rest of the paper we will use these schematic band structure representations to indicate weather a measurement is performed in the regime of region II or I [for example in (d) and (e) respectively]. (b) Schematic representation of the macroscopic Hall bar and the electrical setup used to measure the longitudinal resistance $R_{xx}$ in (c) and the transverse resistivity $\rho_{xy}$ (d,e). (c) Top and backgate voltage dependence of $R_{xx}$ (bias current $I=5~\rm{nA}$). $\rho_{xy}(B_{\perp})$ is measured at each of the locations marked by circles along the lines L and R, shown in (d) and (e) respectively (bias current $I=10~\rm{nA}$).}
\label{fig:delft1}
\end{figure}

The crossover between trivial and topological regimes induced by gate voltage can be clearly seen in magnetoresistance measurements performed on a large Hall bar made from wafer A [Fig.~\ref{fig:delft1}(b)]. The Hall bar width ($20~\rm{\mu m}$) was large compared to  relevant material length scales, and the separation of lateral contacts ($100~\rm{\mu m}$) was much longer than edge scattering lengths in the literature. Positive backgate voltages, $V_{\rm{BG}}$, together with negative topgate voltages, $V_{\rm{TG}}$, raise the electron (conduction) band while lowering the hole (valence) band, creating the band structure of a trivial insulator. When the Fermi energy is tuned into the resulting energy gap, the longitudinal resistance rises to hundreds of $\rm{k\Omega}$ or larger [region \rm{I} in Fig.~\ref{fig:delft1}(c)]. The inverted regime emerges for negative $V_{\rm{BG}}$ and more positive $V_{\rm{TG}}$, that is, when the valence band maximum is driven above the conduction band minimum. When the Fermi energy is tuned into the hybridization gap in the inverted regime, [region \rm{II} in Fig.~\ref{fig:delft1}(c)], the resistance is much smaller compared to region \rm{I}. This is consistent with previous measurements \cite{Qu2015}. Driving the Fermi energy out of the gap, into the conduction (valence) band, yields electron (hole) dominated transport corresponding to regions \rm{III} (\rm{IV}). 

Magnetic field dependence of the transverse resistivity, $\rho_{xy}(B_\perp)$, provides a signature of the gate-induced transition from trivial to inverted band structure \cite{Qu2015}. In the trivial regime, carriers on either side of the charge neutrality point are either purely electron-like or hole-like, giving rise to a $\rho_{xy}$ that is linear in $B_\perp$, in either case \cite{NicheleThesis}. The inverted regime, on the other hand, involves an overlap of electron-like and hole-like carriers near the charge neutrality point, giving rise to  a $\rho_{xy}$ that is non-monotonic in $B_\perp$. Moving the Fermi energy across the gap in the trivial regime [line R in Fig.~\ref{fig:delft1}(c)] yields $\rho_{xy}(B_\perp)$ traces that are linear with slopes passing from negative in the hole regime (point $1$) to positive in the electron regime (point $8$) [Fig.~\ref{fig:delft1}(e)]. At charge neutrality, along line R (point $4$), $\rho_{xy}(B_\perp)$ has large fluctuations but no net slope [Fig.~\ref{fig:delft1}(e)]. A similar set of traces along line L [Fig.~\ref{fig:delft1}(c)], crossing the inverted gap, shows non-monotonic behavior near the charge neutrality point, indicating simultaneous transport of electron- and hole-like carriers [Fig.~\ref{fig:delft1}(d)]. Schematics of the presumed configuration of conduction and valence bands, inverted or noninverted, are shown as figure insets to indicate the regime, topological or trivial, where a particular measurement was carried out [e.g. Fig.~\ref{fig:delft1}(d,e)].

Previous measurements \cite{Qu2015} mapped out the front and back gate dependence of resistivity in the higher-mobility sample C, and correlated features in the zero field resistivity with band structure alignments determined by magnetoresistance. Consistent with the analysis in Ref.~\onlinecite{Qu2015}, the resistivity peak along line L in Fig.~\ref{fig:delft1}(c) marks the crossover from exclusively hole-like transport ($V_{\rm{TG}}\lesssim-2.1~$V) to the overlap region including both conduction and valence bands ($V_{\rm{TG}}\gtrsim-2.1~$V). The lower mobility of sample A, compared to sample C, precludes the observation in Fig.~\ref{fig:delft1}(c) of the resistance peak associated to the charge neutrality point in the inverted regime. Such a feature might appear if the sample were driven farther into the hybridization regime.

In the inverted regime, with the Fermi energy tuned into the hybridization gap (region $\rm{II}$), transport is expected to occur along helical edge channels, and be ballistic over short distances. The edge channels scattering length has been measured in previous work as several microns \cite{Knez2014,Du2015}. Along a $100~\rm{\mu m}$ segment of Hall bar, an edge channel resistance exceeding $h/e^2$ by at least one order of magnitude is therefore expected, whereas the observed resistance peak in region $\rm{II}$ is around $40~\rm{k\Omega}$. This inconsistency may be resolved by including a residual bulk conduction that adds in parallel with the edge channels. One may ask whether a similar explanation is responsible for the residual conductivity in the trivial insulating regime. As we demonstrate below, the answer is no; the finite resistance observed in the trivial insulating regime, Region $\rm{I}$, is instead due to conductive edge channels propagating along the sample perimeter.

\subsection{Non-Local Measurements}
The device geometry described in Fig.~\ref{fig:delft1}(b) measures transport through the bulk in parallel with the edges running between voltage probes. In order to separate bulk and edge contributions, we investigate two measurement geometries: a Hall bar, nominally identical to that of Fig.~\ref{fig:delft1}, measured in a non-local lead configuration, and a Corbino geometry made from the same wafer, where leads are not connected by edges.

\begin{figure}
\includegraphics[width=\columnwidth]{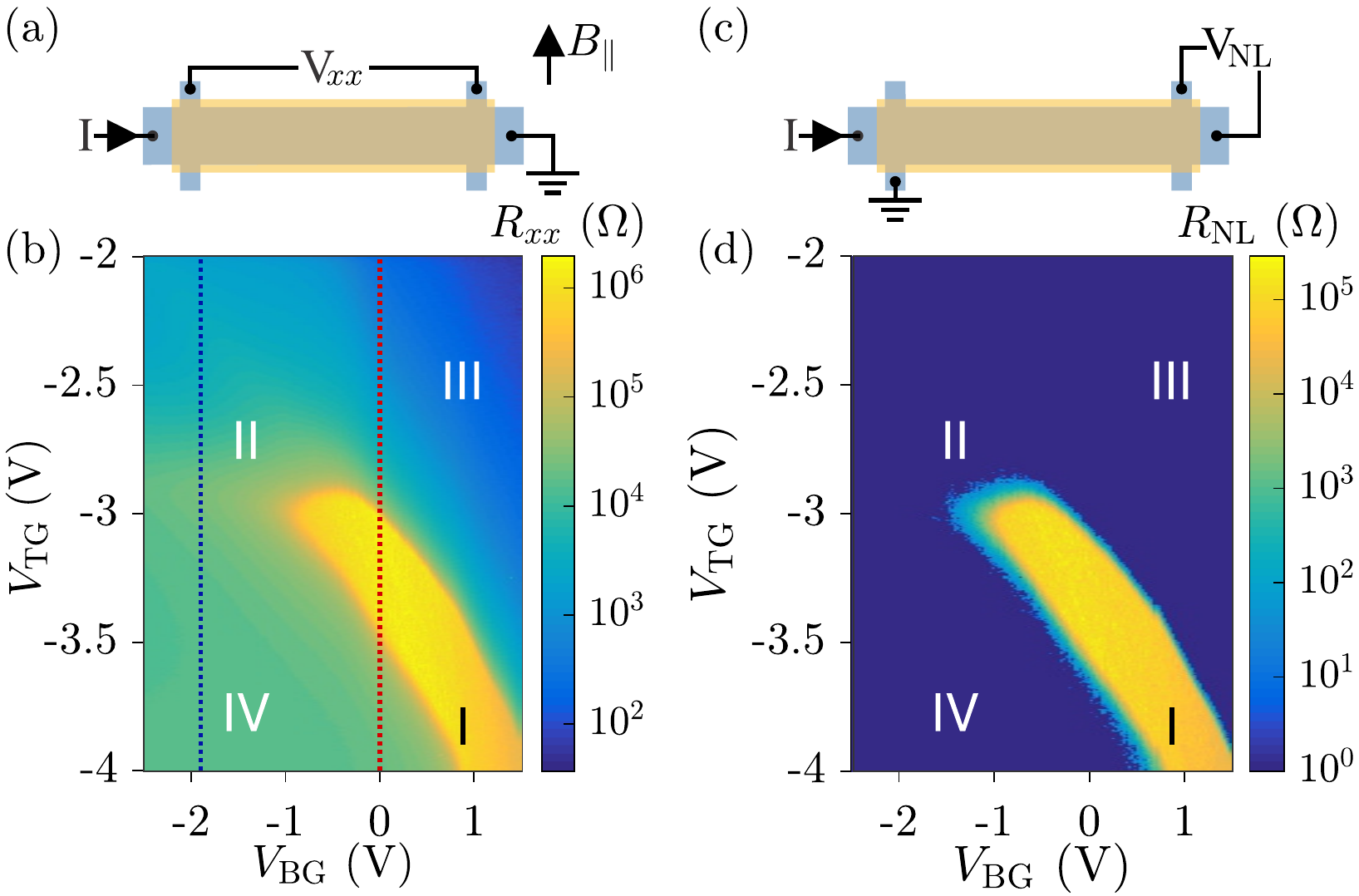}
\caption{(a,c) Schematic representation of the Hall bar geometry and electrical configuration for measuring local longitudinal resistance $R_{xx}$ (b) and non-local resistance $R_{NL}$ (d). The direction of the in-plane magnetic field used in Fig.~\ref{fig:BparVancouver} is indicated. (b) Longitudinal resistance $R_{xx}$ as a function of back gate and top gate voltages. Dotted lines indicate the back gate voltages where temperature [Fig.~\ref{fig:CorbinoCPH}(d)] and in-plane magnetic field (Fig.~\ref{fig:BparVancouver}) measurements were performed. (d) Non-local resistance $R_{NL}$ as a function of back gate and top gate voltages. Note: the color scale in (d) is limited to a minimum of $1~\rm{\Omega}$.}
\label{fig:HBVancouver}
\end{figure}

When current and voltage probes for the Hall bar device geometry from Fig.~\ref{fig:delft1}(b) are rearranged into a non-local configuration, with voltage measured far from the expected bulk current path, the contribution of bulk conduction to the voltage signal will be very small. Quantitatively, the non-local resistance $R_{NL}\equiv dV_{nl}/dI$ due only to diffusive current spreading through the bulk is expected theoretically \cite{Abanin2011} to be suppressed by a factor of $e^{-\pi S}\sim 10^{-7}$ compared to $R_{xx}$, where $S$ is the number of squares between current path and non-local voltage probes. For our device, $S = 5$. On the other hand, edge currents propagating around the sample perimeter would pass the voltage contacts directly and give a sizeable signal.

A comparison of local [Fig.~\ref{fig:HBVancouver}(a,b)] and non-local [Fig.~\ref{fig:HBVancouver}(c,d)] measurements can therefore distinguish bulk-dominated and edge-dominated transport. In particular, the non-local resistance $R_{NL}$ in region $\rm{I}$ is within an order of magnitude of $R_{xx}$, whereas in regions $\rm{II,III,IV}$ $R_{NL}$ is at least four orders of magnitude smaller. Similar measurements for different contact configurations, all around the perimeter of the Hall bar, gave consistent non-local responses. This demonstrates that region $\rm{I}$ is dominated by edge transport, whereas $\rm{II,III,IV}$ are dominated by bulk conduction. Note that the Hall bars in Figs.~\ref{fig:delft1} and ~\ref{fig:HBVancouver} are made from the same wafer, and have the same geometry, but $R_{xx}$ in region $\rm{I}$ is nearly an order of magnitude larger in Fig.~\ref{fig:HBVancouver}(b) compared to Fig.~\ref{fig:delft1}(c). In addition to sample-to-sample variability, this difference may be due to much lower bias currents applied in the insulating region for Fig.~\ref{fig:HBVancouver}(c) measurements ($10~\rm{pA}$) compared to $5~\rm{nA}$ in Fig.~\ref{fig:delft1}(c) \cite{Li2015}.

\subsection{Corbino Disks\label{sec:EdgevsBulk}}
\begin{figure}
\includegraphics[width=\columnwidth]{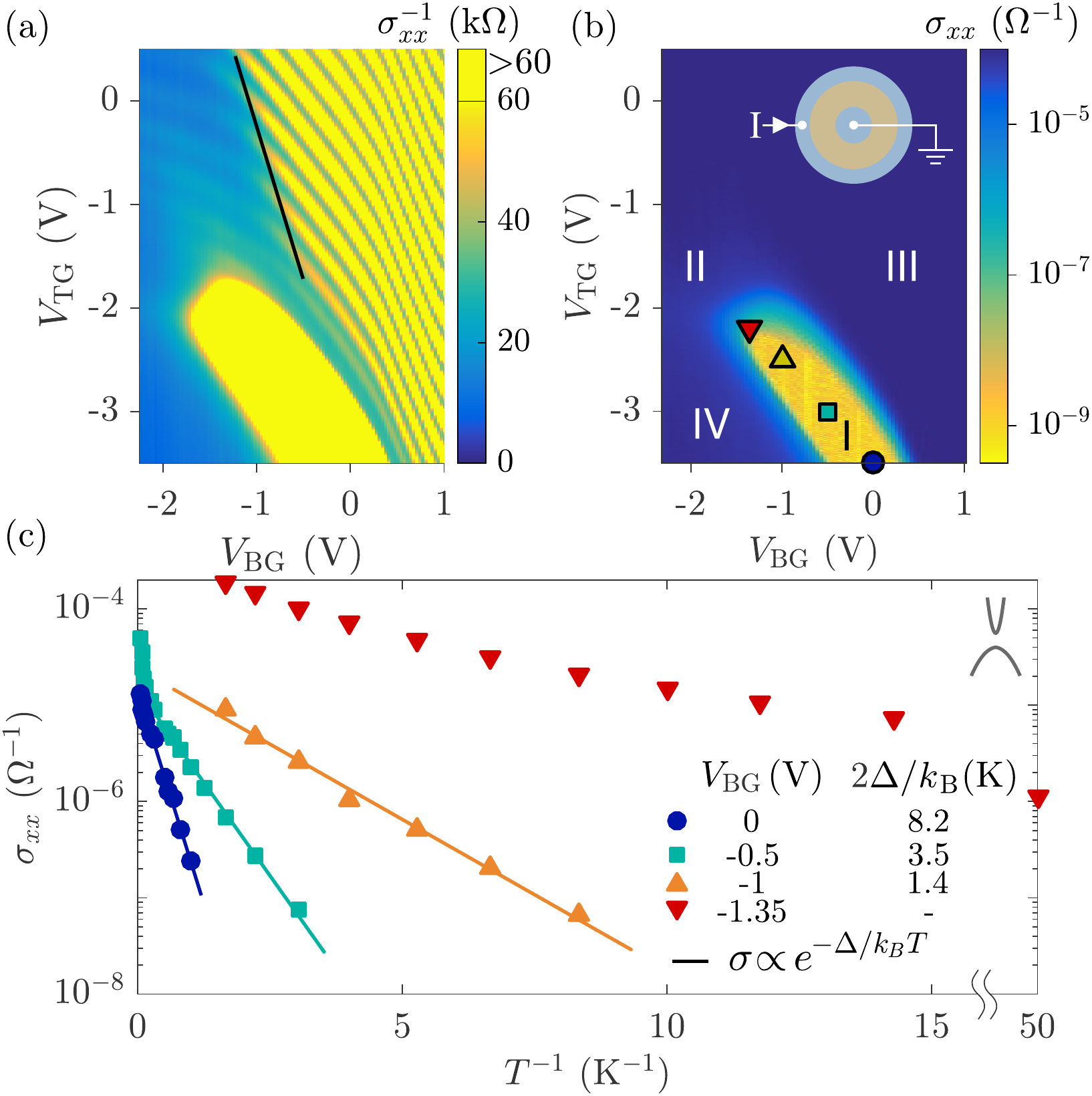}
\caption{(a) Inverse of the longitudinal conductivity, $\sigma^{-1}_{xx}$, of the Corbino disk in an out-of-plane field $B_\perp=1.5~\rm{T}$. The black line marks the slope change in the SdH oscillations, associated to the onset of hole conduction. (b) Longitudinal conductivity $\sigma_{xx}$ measured in a Corbino disk as a function of top gate and back gate voltage. Markers indicate the regimes where the temperature dependence of (c) was taken. Inset: schematic representation of the Corbino geometry. (c) Temperature dependence of the bulk conductivity (markers) together with fits to the Arrhenius law (solid line) at each back gate voltages.}
\label{fig:delft2}
\end{figure}

The non-local measurements presented above indicate that transport in region I is dominated by edge conduction, but do not quantify the degree to which bulk conduction is suppressed ($\sigma_{xx}\rightarrow 0$). To accomplish that, we turn to measurements performed in a Corbino geometry [Fig.~\ref{fig:delft2}(b) inset], in which the current flows exclusively through a ring-shaped bulk separating concentric contacts; no edges connect source to drain. A global top gate overlapping the metallic contacts (but separated by dielectric) tunes the bulk conductance homogeneously. Measurements in Figs.~\ref{fig:delft2} and \ref{fig:CorbinoCPH} were performed on two different Corbino disks in a two-probe configuration. The known series resistance in the cryostat was subtracted from the data.

The first Corbino ring (Fig.~\ref{fig:delft2}), has internal and external radii of $50~\rm{\mu m}$ and $80~\rm{\mu m}$ respectively. In this sample, evidence of the trivial-to-inverted transition is seen in the data of Fig.~\ref{fig:delft2}(a), which shows the inverse of the Corbino conductivity $1/\sigma_{xx}$ at a fixed out-of-plane field of $1.5~\rm{T}$. In the electron regime, clear Shubnikov-de Haas (SdH) oscillations map out contours of constant electron density. Because of the lower mobility and higher effective mass,  SdH oscillations are not visible in the hole regime at the same magnetic field. We interpret the slope change of the SdH oscillations, marked in Fig.~\ref{fig:delft2}(c) with a black line, as the transition from the trivial to the inverted regime. Following the arguments of Ref. \onlinecite{Qu2015}, the coexistence of electrons and holes to the left of the black line results in a decreased back gate capacitance to the electron gas with respect to the right side of the line, where only electrons are present. Similarly, the reduction in the visibility of the oscillations can be attributed to the onset of hole conduction in parallel to the electron system.

When the out-of-plane magnetic field is reduced to zero, the gate voltage map of the conductivity $\sigma_{xx}$ of the Corbino sample [Fig.~\ref{fig:delft2}(b)] looks qualitatively similar to the resistance of the Hall bar [Fig.~\ref{fig:delft1}(c)]. At a quantitative level, however, the resistance of region $\rm{I}$ in the Hall bar is four orders of magnitude lower than the inverse conductivity of the Corbino sample. This can be understood from the fact that the Hall bar geometry in Fig.~\ref{fig:delft1}(b) measures transport via the bulk in parallel with edges that connect $V_{xx}$ voltage probes, whereas the source and drain for the Corbino disk are coupled only via bulk, with no edges. The substantially larger resistance of the Corbino sample therefore indicates that transport in the Hall bar is dominated by conducting edge channels, while the bulk is strongly insulating ($\rm{G\Omega}$ or higher at low temperature).

\subsection{Temperature Dependence}
Bulk conductivity in the trivial regime is strongly temperature dependent. The evolution of the Corbino conductivity, as a function of temperature, extracted for different top gate and back gate voltages from Fig.~\ref{fig:delft2}(b), is shown in Fig.~\ref{fig:delft2}(c). Good agreement with Arrhenius law $\sigma_{xx}\propto\exp(-\Delta/k_{\rm{B}}T)$, with $2\Delta$ the energy gap, over more than two orders of magnitude in resistance [Fig.~\ref{fig:delft2}(c)] indicates activated transport with $2\Delta/k_B$ ranging from $1.4~\rm{K}$ to $8~\rm{K}$. The energy gap increases for more positive back gate voltages, $V_{\rm{BG}}\geq-1~\rm{V}$ [Fig.~\ref{fig:delft2}(c)]. This behavior is qualitatively, but not quantitatively, consistent with a parallel plate capacitor model \cite{Qu2015}, as discussed in Sec.~\ref{sec:discussion}. The temperature dependence for $V_{\rm{BG}}=-1.35~\rm{V}$ is not well fit by an Arrhenius law or a model describing variable range hopping. This is presumably due to the onset of bulk conduction close to the band crossing point.

\begin{figure}
\includegraphics[width=\columnwidth]{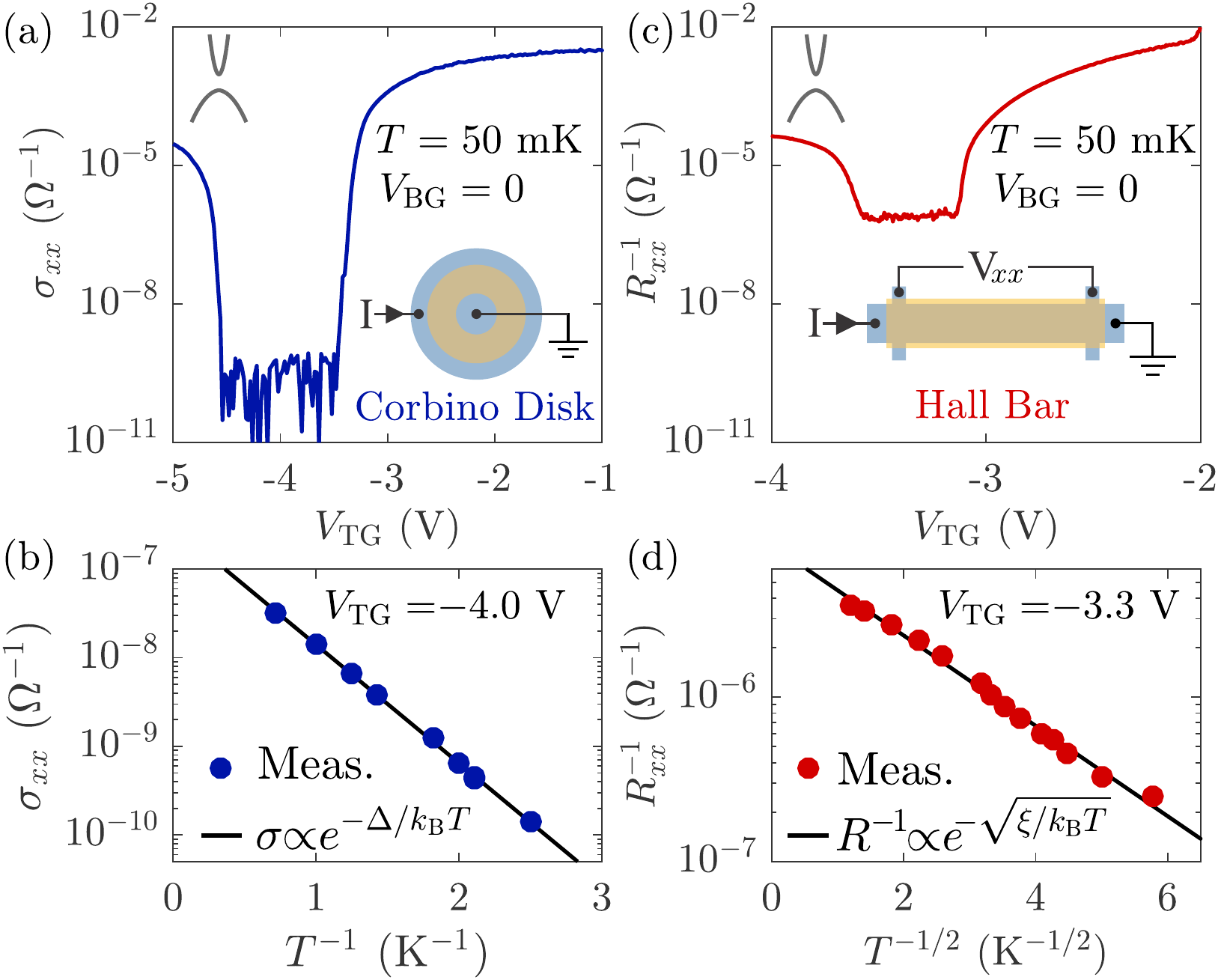}
\caption{(a) Conductivity in the Corbino disk at $V_{\rm{BG}}=0$ as a function of $V_{\rm{TG}}$. Inset: schematic representation of the Corbino disk. (b) Temperature dependence of the conductivity in the Corbino disk (dots) as a function of $T^{-1}$ and a fit to the Arrhenius equation (solid line). (c) Inverse of the longitudinal resistance $R_{xx}$ measured in the Hall bar at $V_{\rm{BG}}=0$ as a function of $V_{\rm{TG}}$. Inset: schematic representation of the measurement setup. (d) Temperature dependence of the Hall bar inverse resistance (dots) together with a fit to the variable range hopping equation (solid line). The horizontal axis is plotted as $T^{-1/2}$ to highlight the consistency with the extracted fit parameter $\alpha=2$.}
\label{fig:CorbinoCPH}
\end{figure}

Similar results were obtained in the second Corbino ring, with internal and external radii of $50~\rm{\mu m}$ and $120~\rm{\mu m}$ respectively. The gate voltage dependence of this device was limited to $V_{\rm{BG}}=0$ due to backgate leakage. Compatibly with the measurement in Fig.~\ref{fig:delft2}(b), the insulating region was characterized by a very low minimum conductivity (experimental noise limited), indicating a strongly insulating bulk [Figs.~\ref{fig:CorbinoCPH}(a)]. A fit to an Arrhenius law [Figs.~\ref{fig:CorbinoCPH}(b)] gives a $V_{\rm{BG}}=0$ energy gap of $2\Delta/k_B=6.2~\rm{K}$, consistent with the previous sample.

Compared to Corbino measurements, the temperature dependence of the conductivity in the Hall bar geometry was much weaker, and inconsistent with the Arrhenius law observed in the bulk [Fig.~\ref{fig:CorbinoCPH}(b)]. Figure~\ref{fig:CorbinoCPH}(d) shows the inverse longitudinal resistance $R_{xx}^{-1}$ measured in the Hall bar of Fig.~\ref{fig:HBVancouver} for $V_{\rm{BG}}=0$ and $T=50~\rm{mK}$. As already noted, the minimum conductance in the Hall bar is four orders of magnitude higher than in the Corbino. Fitting the Hall bar temperature dependence to a more general expression, $R_{xx}\propto\exp(\xi/k_BT)^{1/\alpha}$, with $\xi$ and $\alpha$ as fit parameters, yielded $\alpha=2.0\pm0.5$ compared with $\alpha=1$ for simple activated behavior. The value $\alpha=2$ is consistent with variable range hopping in one dimension or Coulomb dominated hopping in one or two dimensions \cite{Efros1975}. Fixing $\alpha=2$, we obtain $\xi/k_B=(0.4\pm0.04)~\rm{K}$. The insulating temperature dependence of edge resistance observed in these measurements, as well as the strong dependence on bias current or voltage observed at very low temperatures, are qualitatively consistent with recent reports of Luttinger liquid behavior in InAs/GaSb edge modes \cite{Li2015}. However, the data of Fig.~\ref{fig:CorbinoCPH} were not well fit by the specific functional forms used in Ref.~\onlinecite{Li2015}.

\subsection{In-Plane Magnetic Field Dependence}
\begin{figure}
\includegraphics[width=\columnwidth]{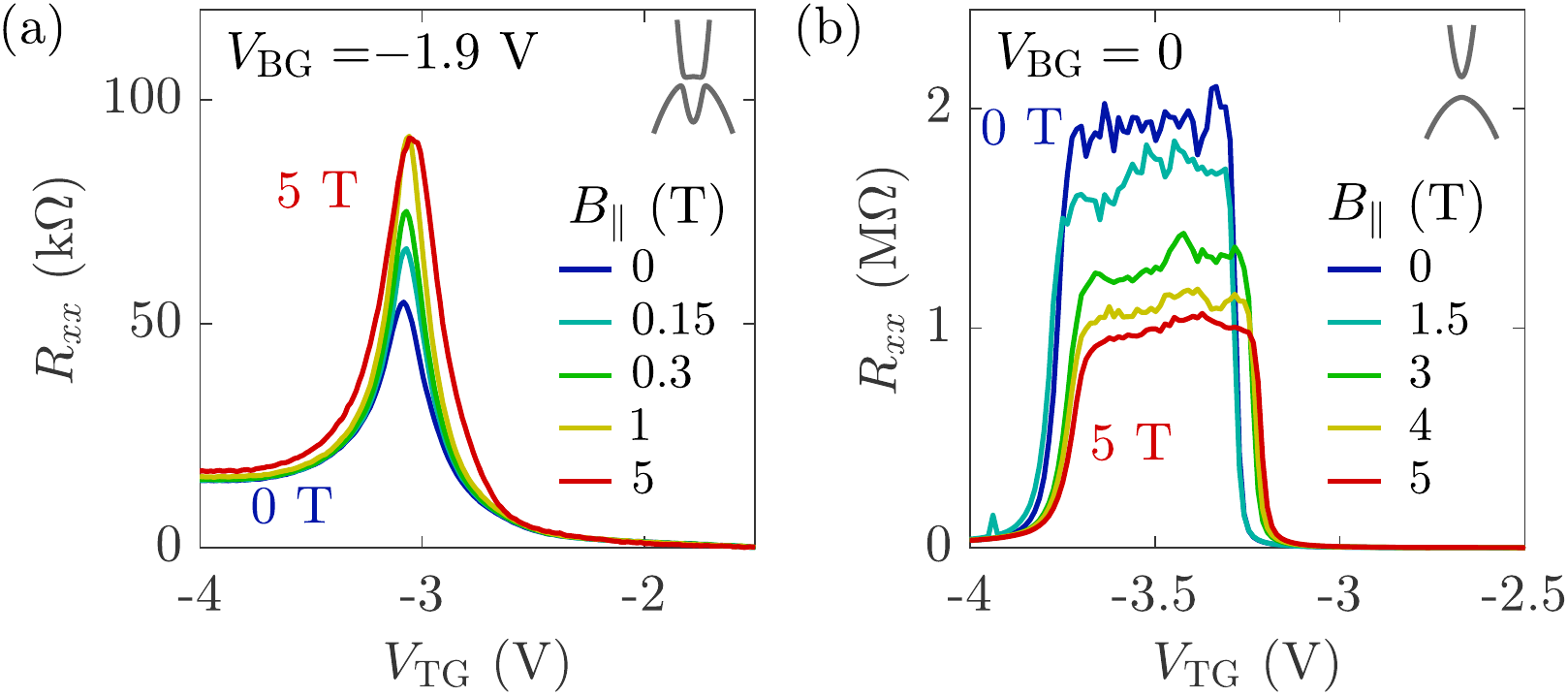}
\caption{(a) Hall bar resistance from Fig.~\ref{fig:HBVancouver}(c) at $V_{\rm{BG}}=-1.9~\rm{V}$ for different values of in-plane magnetic field. The resistance peak is associated to the onset of the conduction band in the hole regime. (b) As in (a), for $V_{\rm{BG}}=0$. The resistance peak marks the trivial gap with edge channel conduction. The field direction in (a) and (b) is indicated in Fig.~\ref{fig:HBVancouver}(a).}
\label{fig:BparVancouver}
\end{figure}
The effect of an in-plane field, $B_\parallel$, on transport in InAs/GaSb in principle provides a means of distinguishing trivial and inverted regimes. The in-plane magnetic field shifts electron and hole bands relative to each other in momentum, quenching the hybridization gap in the inverted regime but leaving the trivial gap largely unaltered \cite{Yang1997,Qu2015}. In the present experiment, however, the quenching of the hybridization gap in the inverted regime cannot be clearly seen due to the large residual bulk conduction that mask the charge neutrality point.
Figure~\ref{fig:BparVancouver}(a) shows the in-plane magnetoresistance for gate voltage settings ($V_{\rm{BG}}=-1.9~\rm{V}$) that give rise to a inverted band alignment. As already noted in Sec.~\ref{magnetotransport}, the resistance peak in Fig.~\ref{fig:BparVancouver}(a) is associated with the onset of the conduction band in the hole regime, not with the charge neutrality point.
The large positive magnetoresistance at low field ($B_\parallel<1~\rm{T}$) cannot be explained simply by quenching of the hybridization gap as described above (it has the wrong sign), but may instead reflect anti-localization for this material, whose bulk resistivity $\rho_{xx}<h/e^2$ places it within the metallic regime. The in-plane field effect saturates above $B_\parallel=1~\rm{T}$, as expected for anti-localization when the Zeeman splitting exceeds relevant spin-orbit energies \cite{Aleiner2001,Meijer2005}.

The weak in-plane field dependence of edge transport in InAs/GaSb in previous experiments \cite{Knez2014, Du2015} remains a difficult aspect of connecting data to a helical edge picture. Similar results were obtained here by measuring the magnetoresistance of the trivial edge channels. In our experiments, only the most resistive device (the Hall bar of Fig.~\ref{fig:HBVancouver}) showed significant in-plane field dependence in the trivial regime: a factor-of-two reduction in resistance at high field in region $\rm{I}$ [Fig.~\ref{fig:BparVancouver}(b)]. The in-plane field dependence was less than $10\%$ for all other devices measured [see, for example, Fig.~\ref{fig:Hbar}(e)]. This  sample-to-sample variability is not yet understood, but is consistent with an origin extrinsic to the edge states themselves. It may also reflect the wide variation of in-plane field dependences observed for variable range hopping that results from a competition between orbital and spin effects \cite{Ioffe2013}. 

\section{Microscopic Samples\label{micro}}

\subsection{Two Terminal Device\label{sec:twoterminal}}

\begin{figure}
\includegraphics[width=\columnwidth]{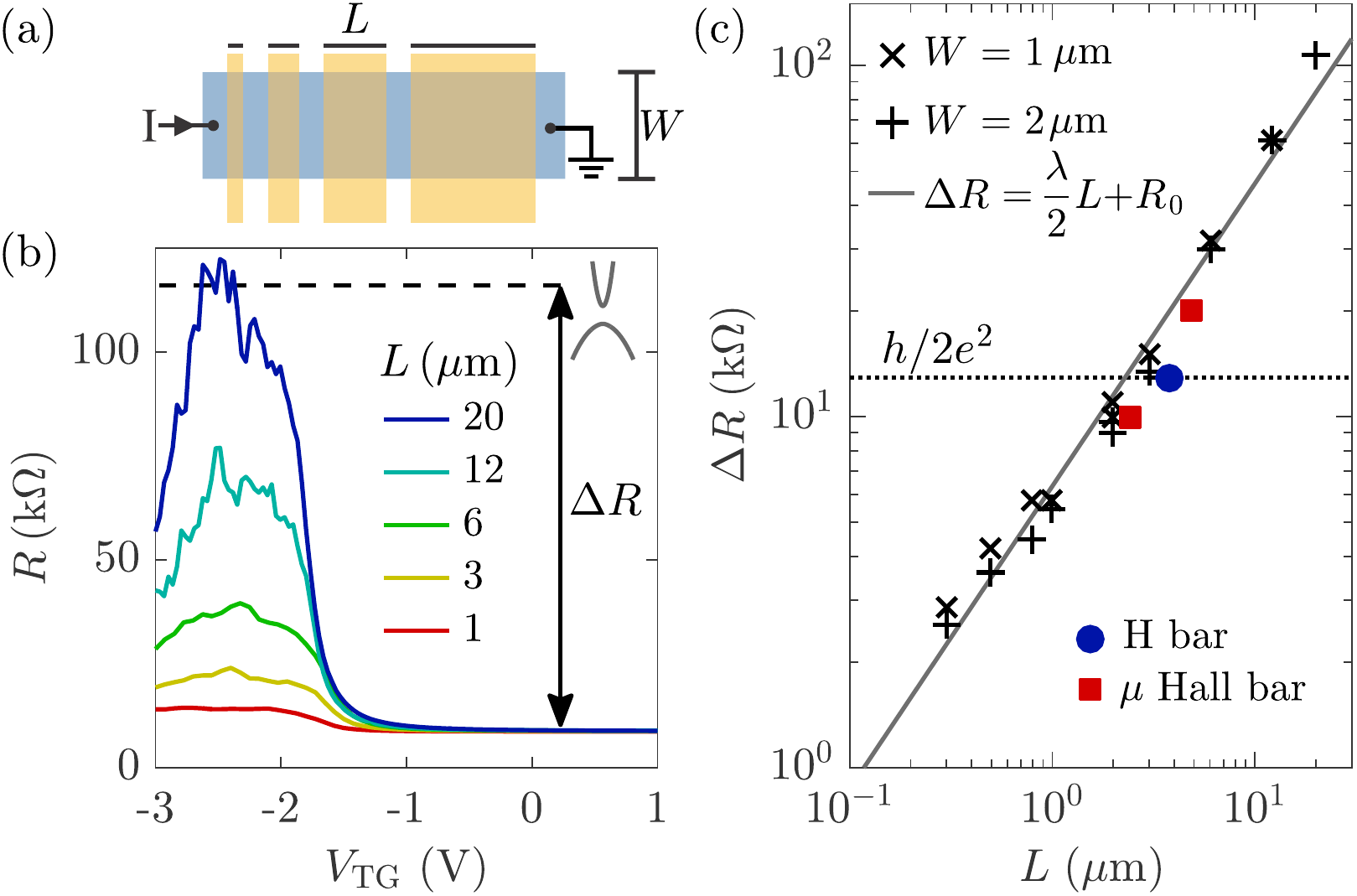}
\caption{(a) Schematic representation of the two-terminal device and the electric setup used to measure the length dependence of the edge channel resistance. (b) Resistance of the $W=2~\rm{\mu m}$ sample as a function of top gate voltage $V_{\rm{TG}}$ for top gates of different lengths $L$. (c). Resistance change in the two-terminal device as a function of gate length for the $W=1~\rm{\mu m}$ (crosses) and $W=2~\rm{\mu m}$ (plus signs) together with a linear fit (black line). Circles and squares indicate the edge resistances measured in the H bar and $\mu$ Hall bar respectively, as discussed in Sec.~\ref{microsamples}.}
\label{fig:LdepCPH}
\end{figure}

Given the similarity between observations of edge transport in our samples, compared to those reported to be in the QSH regime, we next investigate whether the edge channels responsible for the data in Fig.~\ref{fig:HBVancouver} are single-mode, as expected for the spin-resolved edge states of a QSH insulator. Helical edge channels are expected to have quantized conductance, $e^2/h$ for each edge, for edges shorter than a characteristic spin flip length \cite{Bernevig2006,Konig2007,Roth2009,Du2015}. This length has been reported to be several microns in previous work \cite{Du2015}. We tested the quantization of edge channel conductance in our samples using two devices with a geometry similar to that shown in Fig.~\ref{fig:LdepCPH}(a): long InAs/GaSb mesas of width $W$ ($W=1~\rm{\mu m}$ and $W=2~\rm{\mu m}$ for the two devices), across which multiple gates of length $L$ are patterned. The length $L$ of each gated region along the Hall bar ranged from $300~\rm{nm}$ to $20~\rm{\mu m}$.

Starting with all gates grounded, $V_{\rm{BG}}=0$ and either no top gate or all $V_{\rm{TG}}=0$, the entire mesa was in the $n$-doped regime and highly conductive. By monitoring the mesa resistance end-to-end while biasing one gate at a time, bringing the region under the biased top gate into the trivial insulating regime, we determine the edge resistance as a function of length in a single device. 
The effect of various top gates on the two-terminal mesa resistance $R$ is shown in Fig.~\ref{fig:LdepCPH}(b) ($W=2~\rm{\mu m}$). The resistance change $\Delta R$ measured from the resistance peak ($V_{\rm{TG}}\sim-2.2V$) to the highly conductive $n$-type regime ($V_{\rm{TG}}>0V$) represents the resistance of a two-terminal sample with length equivalent to the gate width $L$. The residual length-dependent bulk contribution due to the bulk resistance at positive $V_{\rm{TG}}$ is negligible ($<1\%$) compared to $\Delta R$.

The quantity $\Delta R$ is seen to be directly proportional to $L$ (Fig.~\ref{fig:LdepCPH}c) throughout the range $300~\rm{nm}$ $\leq  L \leq 20~\rm{\mu m}$, and independent of the mesa width $W$ (crosses and plus signs for $W=1~\rm{\mu m}$ and $W=2~\rm{\mu m}$ respectively). The insensitivity of the two-terminal resistance to sample width provides further evidence that current exclusively flows along the edges. The resistance change $\Delta R$ is fit with the functional form $\Delta R=1/2~\lambda L+R_0$ where $\lambda$ is the resistance per unit length of one edge channel, the factor $1/2$ takes into account two edge channels that conduct in parallel, and $R_0$ is the resistance minimum in the short-channel limit. The fit (solid line) results in $\lambda=10.4~\rm{k\Omega\mu m^{-1}}$ and $R_0\approx 0$. This key observation, that transport in the trivial insulating regime is via edge states, with resistance proportional to edge length, will be discussed further Sec.~\ref{sec:discussion}.

\subsection{H Bar and Microscopic Hall Bar\label{microsamples}}
\begin{figure}
\includegraphics[width=\columnwidth]{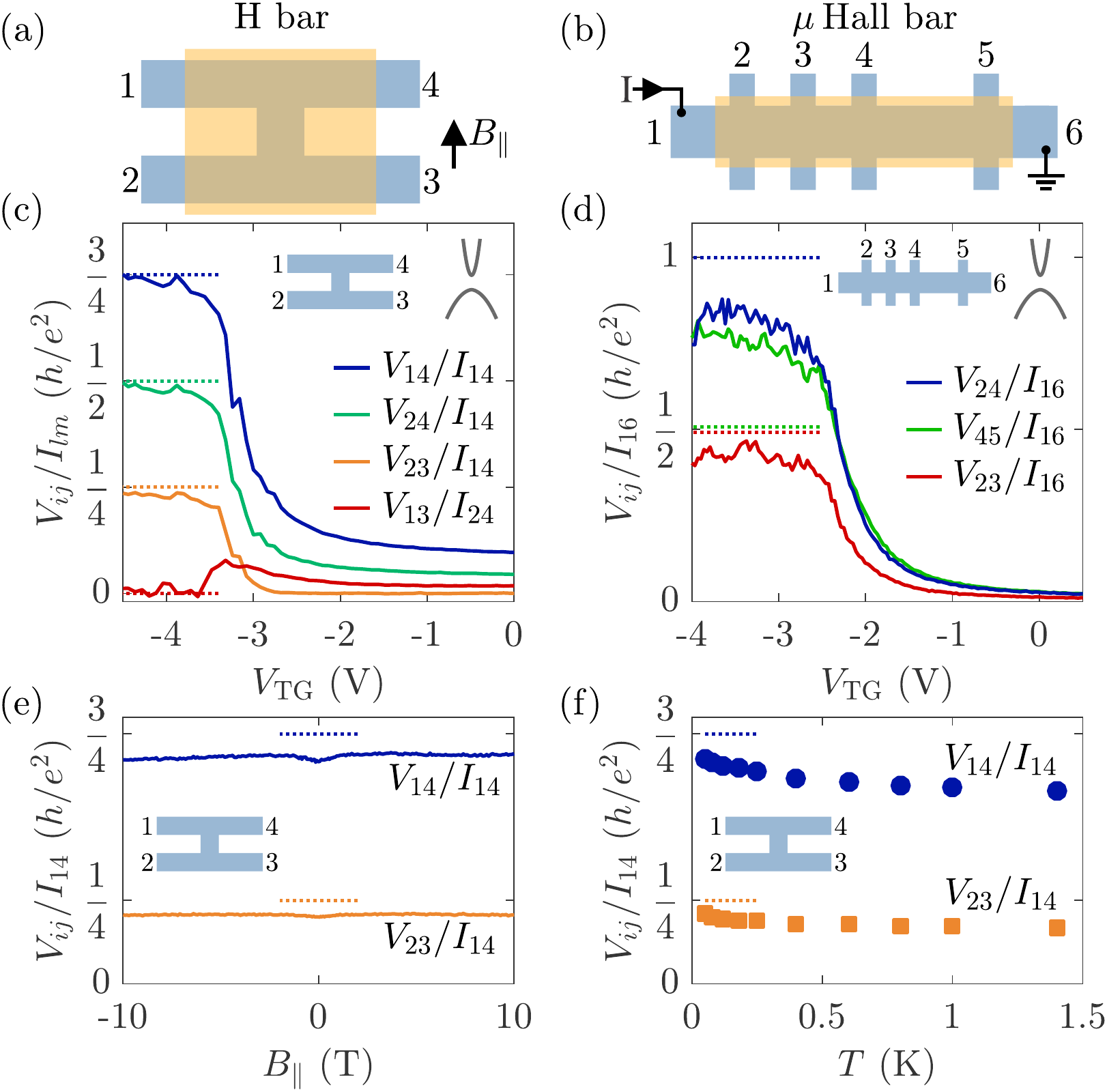}
\caption{(a) Schematic representation of the H bar geometry with the contact numbering used in (c), (e) and (f). (b) Schematic representation of the $\mu$ Hall bar, the electrical setup and the contact numbering used in (d). (c) Four terminal resistances measured in the H bar geometry as a function of top gate voltage in different contact configurations. (d) Four terminal resistances measured in the $\mu$ Hall bar geometry as a function of top gate voltage for different contact configurations. (e) Two H bar four terminal resistances at $V_{\rm{TG}}=-4~\rm{V}$ as a function of in-plane magnetic field. The field orientation is shown in (a). (f) Same as in (e) as function of temperature. Dotted lines in (c), (d), (e) and (f) indicate the expected resistances in case of helical edge channels.}
\label{fig:Hbar}
\end{figure}

One of the strongest arguments in favor of a QSH interpretation for edge channel conduction in previous InAs/GaSb measurements has been the fact that local and non-local resistances of micron-scale structures are close to the quantized values predicted for single-mode edges. The majority of such measurements have been in so-called H bar geometries, or in microscopic Hall bars with micron-scale separations between leads \cite{Konig2007,Roth2009,Knez2014,Du2015}. Notwithstanding the evidence presented above for a non-topological interpretation for edge channel conduction in our samples, we note that characteristic local and non-local resistances for specifically sized (micron-scale) devices in our samples [Figs.~\ref{fig:Hbar}(a) and \ref{fig:Hbar}(b)] were close to values predicted from a Landauer-Buttiker analysis for single-mode edges.

The H bar device, schematically shown in Fig.~\ref{fig:Hbar}(a), has a length of $3.8~\rm{\mu m}$ (defined by the top gate) and arm width of  $1~\rm{\mu m}$. This geometry is nearly identical to those reported in Refs.~\onlinecite{Roth2009,Du2015}. The resulting H shape connects adjacent pairs of $n$-doped contacts by edges each having a length of $3.8~\rm{\mu m}$. Figure~\ref{fig:Hbar}(c) shows various four terminal resistance $R_{ij-lm} = V_{ij}/I_{lm}$, measured by passing a current $I_{lm}$ between terminal $l$ and $m$ and by measuring the voltage drop $V_{ij}$ between contact $i$ and $j$. When the top gate drives the bulk into the insulating regime ($V_{\rm{TG}}<-3~\rm{V}$), the resistance saturates to a plateau that depends on the particular set of contacts used for the measurement.

The plateau resistances are very close to the quantized values expected in this geometry for perfectly transmitting helical (single-mode) edge channels, as calculated using Landauer-B\"uttiker formalism \cite{Buettiker1988,Roth2009}. The configuration $V_{14}/I_{14}$ (blue line) can be modeled as one $h/e^2$ resistor (direct path from $1$ to $4$) in parallel with three $h/e^2$ resistors in series, yielding a total resistance of $3h/4e^2$. Similar arguments hold for the other three curves shown in Fig.~\ref{fig:Hbar}(c). The four terminal resistance $V_{23}/I_{14}$ (orange line) measures exclusively the non-local response of the edge channel. When the sample is in the $n$-type regime (i.e. for $V_{\rm{TG}}>-3~\rm{V}$), $V_{23}/I_{14}$ vanishes. For $V_{\rm{TG}}<-3~\rm{V}$, a plateau at $h/4e^2$ forms offering further evidence that, also in the H bar, transport in the insulating regime is exclusively mediated by edge channels. The symmetric configuration $V_{13}/I_{24}$ (red line) results in a zero resistance plateau. The zero resistance plateau indicates that the currents moving on opposite sides of the H bar are balanced. We stress that this analysis assumes single mode channels, which the length dependent measurements presented in the previous section appear to rule out. We are therefore left to interpret this apparent resistance quantization [Fig.~\ref{fig:Hbar}(c)] as coincidental, due to an edge channel resistance of roughly $26~\rm{k\Omega}$ $\approx h/e^2$ for these particular device sizes.

The same type of analysis is performed on a sample with a more conventional Hall bar geometry where the separation between adjacent contacts is on the micron scale, shorter than previously reported relaxation lengths, referred to as a $\mu$-Hall bar. As shown in Fig.~\ref{fig:Hbar}(b), the $\mu$-Hall bar has eight lateral arms, a width of $1~\rm{\mu m}$ and a length of $12~\rm{\mu m}$ (defined by the top gate). The separation between contact $2$ and $3$ and between $3$ and $4$ is $2.4~\rm{\mu m}$; the separation between contact $4$ and $5$ is $4.8~\rm{\mu m}$. The sample is measured by passing a current $I_{16}$ between contact $1$ and $6$ and measuring the voltage drop between pairs of lateral arms. For the case of perfectly transmitting helical edge channels, the four terminal resistance would be $h/2e^2$ if measured between adjacent lateral arms, independent of spatial separation. If measured between two lateral arms separated by a third arm acting as a dephasing probe, the four terminal resistance would rise to $h/e^2$, the classical resistors-in-series result. Contrary to these expectations for quantized edges, the measured resistance depends exclusively on length, and is not dependent on the number of intervening voltage probe contacts, as shown in Fig.~\ref{fig:Hbar}(d). In particular, $V_{23}/I_{16}$ (red line, $2.4~\rm{\mu m}$) is half of $V_{45}/I_{16}$ (green line, $4.8~\rm{\mu m}$), while they should both be quantized at $h/2e^2$. Similarly, the presence of an unused voltage probe between contacts 2 and 4 does not elevate the resistance $V_{24}/I_{16}$ (blue line,$4.8~\rm{\mu m}$) above $V_{45}/I_{16}$ (green line, $4.8~\rm{\mu m}$), for which there is no voltage probe between contacts.

The local and non-local resistances of the H bar were found to depend much less strongly on temperature or in-plane field as compared to analogous measurements in the macroscopic Hall bar. Figure~\ref{fig:Hbar}(e) and (f) show the $V_{14}/I_{14}$ and $V_{23}/I_{14}$ configurations for a constant top gate voltage $V_{\rm{TG}}=-4~\rm{V}$ as a function of $B_\parallel$ and $T$ respectively. The resistance of the edge channels does not show any significant field dependence up to $10~\rm{T}$, except a weak positive magnetoresistance close to $B_\parallel=0$. We interpret these observations as consistent with the evolution from a temperature independent regime above $0.5~\rm{K}$ to a weakly insulating dependence approaching low temperatures.

The samples presented in this section could not be operated at finite back gate voltage due to the onset of leakage currents. Even if the inverted regime could be reached in these samples, it would be difficult to detect the edge channels by conventional transport methods due to the presence of large bulk conduction, as presented in Sec.~\ref{macro}.

\section{Scanning SQUID Measurements\label{squid}}

\begin{figure}
\includegraphics[width=\columnwidth]{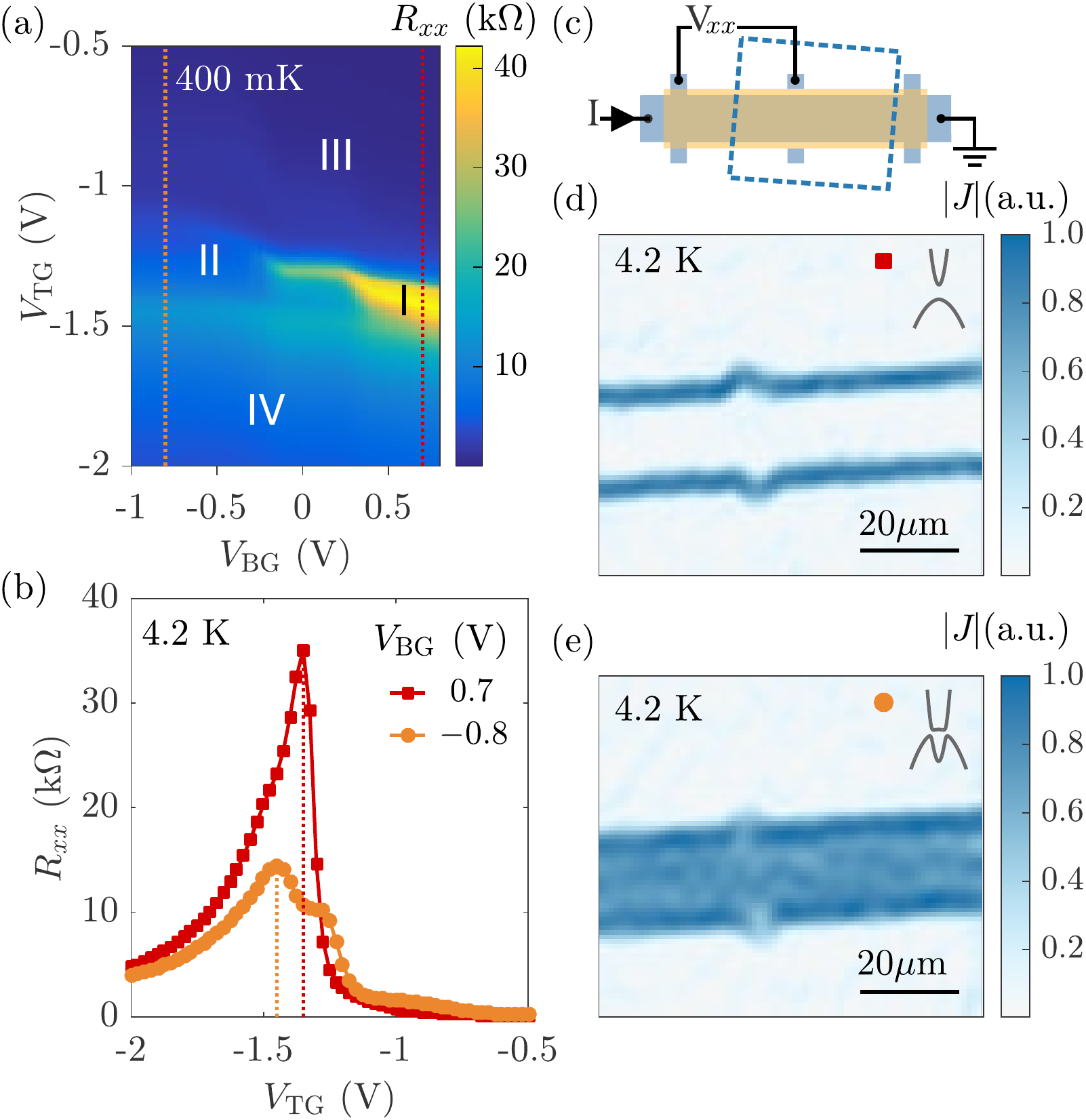}
\caption{(a) Four-terminal longitudinal resistance $R_{xx}$ as a function of top gate ($V_{\rm{TG}}$) and back gate voltage ($V_{\rm{BG}}$). For clarity, a higher resolution 2D plot taken at $400~\rm{mK}$ is shown in lieu of one taken at $4.2~\rm{K}$, the same temperature as the images. (b) Resistance traces vs. $V_{\rm{TG}}$ taken at $4.2~\rm{K}$ for different back gate voltages $V_{\rm{BG}}$, as indicated by the dotted lines in (a). (c) Schematic representation of the measurement setup. The dashed box indicates the imaged area. (d,e) Scanning SQUID images of the absolute value of the current density $|J|$, acquired in the high resistance non-inverted regime (d) and lower resistance, inverted regime (e). The images were taken at back gate voltages indicated in (b) by the dashed lines.}
\label{fig:SQUID}
\end{figure}

To complement our investigation of edge conduction via transport measurements, we next present results of direct spatial imaging of edge conduction using scanning SQUID microscopy. The measurements were performed on a Hall bar of equal dimension as in Figs.~\ref{fig:delft1} and \ref{fig:HBVancouver} obtained from Wafer C. The SQUID used to image current had a $3~\rm{\mu m}$ diameter pickup loop \cite{Huber2008}. An alternating current was applied to the sample [Fig.~\ref{fig:SQUID}(c)] and the AC flux response was measured through the SQUID's pickup loop as a function of position. Using Fourier techniques and our SQUID experimentally-extracted point spread function \cite{Roth2009,Nowack2013}, the 2D current density was obtained directly from AC flux images. The images in Fig.~\ref{fig:SQUID} present the absolute value of the 2D current density, which in this geometry is roughly proportional to the local conductivity. Current density images were taken at $500~\rm{nA_{rms}}$, which is rather high compared to currents used in standard transport measurements. The shape of flux line cuts in the trivial regime did not change as a function of applied current, down to $50~\rm{nA_{rms}}$. The relatively high bias (up to $10~\rm{mV}$ across the voltage probes) of these measurements most likely masks any non-linear effects present at lower biases \cite{Li2015}. For this experiment, unintentional RC filtering from the wiring had not been well characterized at the frequencies of the applied current, so the extracted current density images are plotted in arbitrary units (A.U.). Transport measurements on the device imaged by SQUID were taken at $10~\rm{nA_{rms}}$ and low frequencies ($\sim 10~\rm{Hz}$), using the contacts indicated in Fig.~\ref{fig:SQUID}(c).

\begin{figure}
\includegraphics[width=\columnwidth]{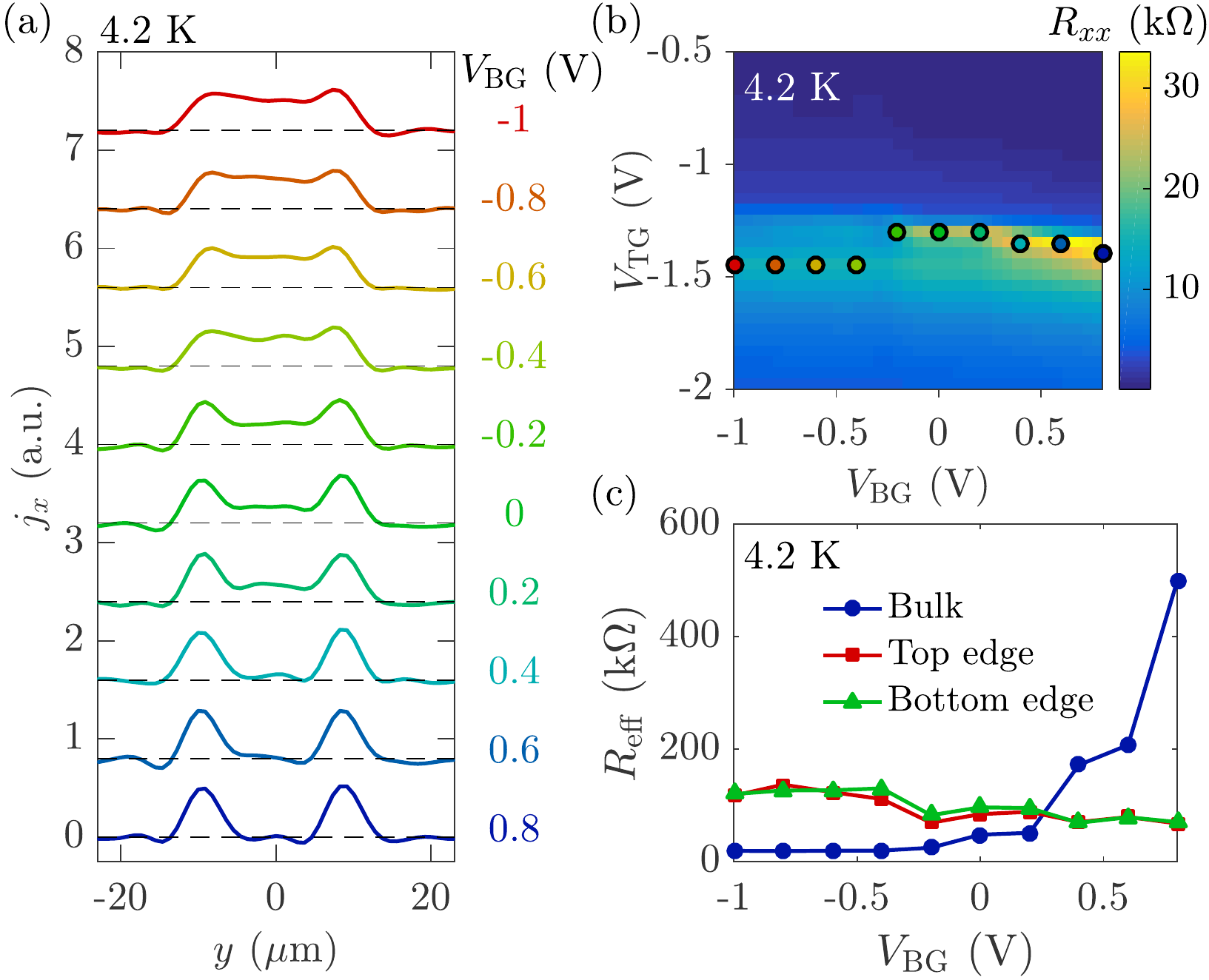}
\caption{(a) Line cuts of current density $j_x$ extracted from averaged flux line cuts as a function of back gate voltage. The line cuts are offset for clarity. Each line cut was taken at the resistance maximum of the top gate sweep, which is not necessarily the charge neutrality point in the inverted regime. The applied current was $I=100~\rm{nA_{rms}}$. (b) Longitudinal resistance taken at $T=4.2~\rm{K}$. The gate voltages at which the line cuts of (a) were taken are indicated by the corresponding markers. (c) Effective resistance $R_{\rm{eff}}$ extracted from fitting flux line cuts and the measured $R_{xx}$.}
\label{fig:SQUID2}
\end{figure}

The gate voltage map of resistance for wafer C [Fig.~\ref{fig:SQUID}(a)] was qualitatively but not quantitatively similar to the analogous maps for wafer A previously presented in Figs.~\ref{fig:delft1}, \ref{fig:HBVancouver}, and \ref{fig:delft2}. Thorough magnetotransport studies of wafer C from Ref.~\onlinecite{Qu2015}, covering similar gate voltage ranges, confirms the labeling of the phase diagram into regions $\rm{I, II, III, IV}$ as in Fig.~\ref{fig:delft1}. Resistance peaks as a function of top gate voltage [Fig.~\ref{fig:SQUID}(b)] identify the alignment of the Fermi energy within the inverted and trivial gaps. In the trivial regime, the resistance rises only to $35~\rm{k\Omega}$ [Fig.~\ref{fig:SQUID}(b)] compared to hundreds of k$\Omega$ observed in wafer A. In the inverted regime, the resistance peak is around $R_{xx} \sim 15~\rm{k\Omega}$ compared to $40~\rm{k\Omega}$ in wafer A. Note that the 2D resistance plot in Fig.~\ref{fig:SQUID}(a) was taken at $400~\rm{mK}$, whereas the scanning images [Fig.~\ref{fig:SQUID}(d,e)] were measured at $4.2~\rm{K}$. The transport data at $4.2~\rm{K}$ is qualitatively similar [see Fig.~\ref{fig:SQUID2}(b)], although the resistance peak in the trivial regime is lower.

The main scanning SQUID results are presented in Figs.~\ref{fig:SQUID}(d,e). In Fig.~\ref{fig:SQUID} images were taken far from the point of band gap closing, near the largest positive and negative values of $V_{\rm{BG}}$ applied. Specifically, current was imaged at the maximum resistance at fixed back gate voltages $V_{\rm{BG}}=0.7~\rm{V}$ (trivial) and $V_{\rm{BG}} = -0.8~\rm{V}$ (inverted), indicated by the dashed red and yellow lines in Fig.~\ref{fig:SQUID}(a). In the trivial regime, current flowed exclusively on the edge of the sample [Fig.~\ref{fig:SQUID}(c)], consistent with the conclusion reached from the transport data presented above. Even at the small overlap between the top-gate and the voltage leads [see Fig.~\ref{fig:SQUID}(c)] current flows along the edge of voltage probes until it reaches the ungated $n$-type region. When the Fermi energy was far from the gap, in either the $n$-type or $p$-type conducting regimes III or IV of the phase diagram, no edge currents were observed (not shown).

In the inverted regime ($V_{\rm{BG}}=-0.8~\rm{V}$), enhanced current density along the edges of the device was also observed, but concomitantly to measurable current flow in the bulk. This is consistent with what was observed previously in undoped InAs/GaSb with scanning SQUID \cite{Spanton2014}, and explained by the residual bulk conductivity in the hybridization gap, as confirmed by Corbino measurements [Figs.~\ref{fig:delft2} and~\ref{fig:CorbinoCPH}].
The measurements in the inverted regime were performed at the maximum resistance, which in this case might coincide with the onset of electron-hole hybridization, as in Fig.~\ref{fig:BparVancouver}(a), rather than the middle of the gap. The presence of edge channels throughout the gap has been established by scanning SQUID previously in InAs/GaSb \cite{Spanton2014}, and therefore the presence of edge channels here, even if the Fermi level is not well-centered in the hybridization gap, is not surprising.

Conducting edges were observed across the phase diagram in the gapped regions at all of the back gate voltages which were investigated [see Fig.~\ref{fig:SQUID2}]. For the chosen values of $V_{\rm{BG}}$, $V_{\rm{TG}}$ was set in order to maximize the value of $R_{xx}$ and then the magnetic flux from the Hall bar was imaged along a line perpendicular to the current flow (`flux line cuts'). The positions in gate space where the flux line cuts were taken are indicated on the resistance color plot of Fig.~\ref{fig:SQUID2}(b). The flux line cuts were converted into current density along the Hall bar axis, $j_x$, using methods described elsewhere \cite{Spanton2014}. The result of this analysis is shown in Fig.~\ref{fig:SQUID2}(a). Edge states were present throughout the entire phase diagram (on the resistance peaks), and the current along the edges and in the center of the device varied smoothly as a function of $V_{\rm{BG}}$. The bulk current was nearly zero in the non-inverted regime ($V_{\rm{BG}}=0.8~\rm{V}$), and rose smoothly above zero as the gate voltage was tuned into the inverted regime.

In order to quantify the dependence of the current distribution on $V_{\rm{BG}}$, the flux line cuts of Fig.~\ref{fig:SQUID2}(a) were fitted to determine the fraction of current $F$ flowing in the top edge, bottom edge and bulk of the Hall bar ($F_{\rm{top}}$, $F_{\rm{bot}}$, $F_{\rm{bulk}}$). Assuming each of the three channels contributes to transport in parallel, their effective resistances are given by $R_{\rm{eff}}=R_{xx}/F_{\rm{top, bot, bulk}}$ (for more details see Ref.~\onlinecite{Spanton2014}). The results of this analysis are shown in Fig.~\ref{fig:SQUID2}(c). Consistent with previous observations, the effective resistance of the bulk strongly increased at positive $V_{\rm{BG}}$, indicating an opening of the trivial gap. On the other hand, in the inverted regime we found that the bulk effective resistance does not change significantly as a function of $V_{\rm{BG}}$. Additionally, there is not a strong decrease in the bulk resistance in between the two regimes, as one would expect for a gap closing. Both of these features are at least partially explainable by the residual bulk conductivity in the inverted regime. Despite the transition from inverted to non-inverted regime, the edges effective resistance varied only up to a factor of two between the highest and lowest $V_{\rm{BG}}$ values. In particular, the edges resistance smoothly changed across the region where the gap should close, consistent with the edges observed in the non-inverted regime persisting into the inverted regime. It is possible, however, that the similarity in the resistance of the edges in the two regimes is accidental, and that the trivial edge states disappear only close to the gap closing. More detailed work near the gap closing is warranted, especially at lower biases and temperature, but these measurements indicate that the presence of trivial edge states in the inverted regime, in addition to the trivial regime, is certainly possible.

\section{Discussion\label{sec:discussion}}
The ability to tune between inverted and trivial regimes using top and back gate voltages enables a determination of the sample band structure topology for a given set of conditions \cite{Qu2015}. As outlined above, however, we observe several surprising characteristics of the trivial phase for this sample. 

First, the temperature dependence of the bulk conductivity measured in Corbino geometry implies an energy gap $\Delta\leq8\rm{K}$ that is surprisingly small compared to theoretical expectations \cite{Yang1997,Liu2008}. Using a parallel plate capacitor model \cite{Qu2015}, the estimated energy gap at $V_{\rm{BG}}=0$ would be $300~\rm{K}$ assuming that electron and hole wavefunctions sit in the center of the respective QWs, and that the gap closes when $V_{\rm{BG}}=-1~\rm{V}$ [this backgate voltage corresponds to the tip of the trivial phase in Fig. \ref{fig:delft2}(a)]. The two orders of magnitude discrepancy between the measured and estimated energy gap in the trivial regime is not understood. The electron and hole wavefunctions separation could be much smaller than the QWs thickness or, as observed in bilayer graphene, disorder may result in a large underestimate of the energy gap size measured in a transport experiment \cite{Taychatanapat2010,Min2011}.

Most significantly, edge channels are consistently observed in the trivial regime, both in transport and in scanning SQUID images. The non-topological character of these edges is supported by measurements indicating that the edge channel resistance scales linearly with length down to a length of at least $300~\rm{nm}$, at which point the resistance is far below the $h/2e^2\sim 13~\rm{k\Omega}$, expected for single-mode conducting channels [Fig.~\ref{fig:LdepCPH}(c)]. The resistance of a true QSH sample can increase above $h/2e^2$ in case of spin scattering between counterpropagating edges, but it can not assume lower values (assuming no bulk conduction). Ballistic, single-mode non-helical edge channels would yield a minimum resistance $h/4e^2\sim 6.4~\rm{k\Omega}$ for lengths less than or comparable to the elastic mean free path. With a minimum measured $\Delta R\sim 3~\rm{k \Omega}$ that falls well below this lower bound, we conclude that our edge channels are composed of at least $2$ spin-degenerate modes, with a mean free path shorter than $300~\rm{nm}$.

One of the primary points to be taken from this work is that, following standard recipes, trivial edge modes are likely to be found in InAs/GaSb QW systems conducting in parallel with any helical edge modes that might appear in inverted band structure regime. These modes are consistently observed in the conventional insulating state, and should likely be present in the inverted regime too where (hybridization) band gaps are much smaller. It is worth noting that our observation of edge channels in the trivial regime does not exclude the possibility of finding helical edges in the inverted regime, but in the present samples bulk conduction is too high for these to be observed in a transport experiment. On the other hand, scanning SQUID images offer evidence of enhanced edge conduction in the inverted regime that continuously evolves to edge states in the trivial regime.

The non-helical edge conduction we report is robust in the sense that it was observed for many different samples made on three different wafers and processed in three different laboratories, using different top gate insulators and slightly different processing recipes. This indicates that edge conduction may be a common feature of InAs/GaSb quantum wells. At the same time, quantitative details of the edge conductance did appear to depend on precise processing conditions. For example, the linear edge resistivity $\lambda$ was identical for the two two-terminal devices of Sec.~\ref{sec:twoterminal}: $\lambda=10.4~\rm{k\Omega\mu m^{-1}}$ at $T=50~\rm{mK}$. These devices were patterned on the same chip and processed at the same time. The macroscopic Hall bar described in Fig.~\ref{fig:HBVancouver} was fabricated on a different chip from the same wafer, and processed in a different fabrication run using identical parameters; the linear edge resistivity of this device was $\lambda=26.4~\rm{k\Omega\mu m^{-1}}$. The H bar and the $\mu$-Hall bar, patterned together on a third chip from the same wafer, gave $\lambda=8~\rm{k\Omega\mu m^{-1}}$ (circles and squares in Fig.~\ref{fig:LdepCPH} for the H and $\mu$-Hall bar respectively). 

The temperature dependence and in-plane field dependence of $\lambda$ was also different for samples processed in different batches. Edge channels showed an insulating temperature dependence ($\partial\lambda/\partial T<0$) in every sample, but the magnitude of the variation with temperature was much stronger in the macroscopic Hall bar [see Fig.~\ref{fig:CorbinoCPH}(d)].
The in-plane field dependence of the macroscopic Hall bar was also much stronger: a factor of two resistance decrease in a $5~\rm{T}$ field [see Fig.~\ref{fig:BparVancouver}(b)], compared to a $<5\%$ change for the H bar [see Fig.~\ref{fig:Hbar}] and the two terminal devices. The general magnetic field dependence is not consistent with QSH edge channels, where the breaking of time reversal symmetry is expected to induce back scattering.

In the following we propose different scenarios that could give rise to the observed effects and mention possible solutions. We anyway stress that understanding the origin of the trivial edge channel conduction, and eventually suppressing it, goes beyond the scope of this report.
The processing-dependent linear resistance of the edge channels in these devices may give a hint to their origin. For example, band bending of the InAs conduction band at the vacuum interface can depend on the precise termination of the semiconductor crystal. This effect is typically of the order of the bulk InAs energy gap \cite{Tsui1970,Noguchi1991,Olsson1996}. Because of the relatively small energy gap in the double QW system close to the inverted-trivial transition, band bending can be particularly relevant, leading to a significant charge accumulation at the etched edge of the samples.
While our observations do not preclude the existence of a topological phase in the inverted regime of our samples (region II), observing the effects of true helical edge-channel transport would require controlling the band bending of both electrons and holes to values smaller than the bulk hybridization gap.

Alternatively to band structure effects, spurious effects of the fabrication process might constitute the most relevant contribution to the creation of edge channels. As an example, the side walls of the mesa might become conducting due to a redeposition of amorphous Sb during AlSb etching, or to dangling bonds resulting from the exposure of the etched semiconductor to air \cite{Gatzke1998, Chaghi2009}. Such problems have been widely studied in the field of optoelectronics, and various passivation techniques were proposed \cite{Plis2013}. 

We note that band bending at the sample edges is a phenomenon that has been observed for other small band gap materials. Graphene, for example, exhibits enhanced edge conduction close to the charge neutrality point, as was observed via superconductive interferometry measurements \cite{Allen2016}. It was also recently demonstrated that inverted HgTe/HgCdTe QWs also show edge channels whose conductance properties are inconsistent with the common expectations of a QSH insulator \cite{Ma2015}. The authors of Ref.~\onlinecite{Ma2015} also speculated that, in the case of HgTe/HgCdTe, extrinsic effects may cause an enhanced conductance close to the sample edges. 

Enhanced conductance can also arise due to electric field focusing at the sample edges \cite{Vera-Marun2013}. This effect may be particularly relevant for top gates deposited after etching, resulting in conformal coverage of the etched walls. Because of the higher top gate capacitance at the mesa walls, the edges can be brought to a conductive $p$-type regime for a top gate voltage at which the bulk is still insulating.

If the sample edges have finite carrier density due to band bending or other effects in the trivial regime, one might consider depleting them using side gates. Scenarios for band-banding in InAs/GaSb, and how it can be corrected using additional gates are discussed in Ref.~\onlinecite{Beukman2016}. Preliminary results indicate that side-gating does indeed reduces edge conduction, but not eliminate it.

Recent measurements of Si-doped InAs/GaSb QWs in other groups have confirmed the coexistence of an insulating bulk with conductive edge channels \cite{Knez2014,Spanton2014,Du2015}. Similar to the measurements reported here, the resistance of the edge channels scaled linearly with length, with $\lambda\approx 6~\rm{k\Omega\mu m^{-1}}$. The samples presented in Ref.~\onlinecite{Knez2014,Spanton2014,Du2015} were claimed to be in the inverted regime, whereas the measurements reported here are for samples whose regime (inverted or trivial) can be changed using gate voltages. The most significant contrast between earlier reports \cite{Du2015} and the measurements reported here is the observation of conductance quantization to within $1\%$ of the expected value for three devices with edge lengths somewhat shorter than the typical scattering length scale $\lambda_\varphi\approx4.4~\rm{\mu m}$ \cite{Du2015}. The more extensive measurement of length dependence reported here, extending down to lengths an order of magnitude shorter than $\lambda_\varphi$, enabled a clear determination that in our samples the apparent quantization of edge resistance was coincidental, depending on sample size. 

\section{Conclusion}
We have shown that edge channel transport in InAs/GaSb, previously regarded in the literature as a signature of helical states, is also found in the trivial (non-topological) regime. Quantitative metrics of the edge transport in our samples, with non-inverted band structure, are nearly identical to those described in earlier reports. Two experimental observations, however, allow us to conclude that the edge conduction reported here is of a different nature than that predicted in the framework of the QSH effect: First, we explore the entire phase diagram of our samples via gate voltages, and thereby identify the parameter space where edge conduction is observed to be one where the band structure is trivial, that is, not inverted. Second, short edge channels segments have a resistance much smaller than $h/e^2$, indicating they are composed of many modes with a short scattering length.

Our results highlight the importance of considering enhanced edge conduction in broken-gap materials, where the energy gap might be comparable to band bending at an interface. Trivial edges result in a behavior strikingly similar to those expected for a QSH insulator, hence proper characterization of the edge channels nature is crucial. Our measurements and analysis provide one example of an experimental framework for distinguishing between trivial and helical edge states.

\begin{acknowledgments}
This work was supported from Microsoft Corporation Station Q. The work at Copenhagen was also supported by the Danish National Research Foundation and Villum Foundation. The work at Delft was also supported by funding from the Netherlands Foundation for Fundamental Research on Matter (FOM). The work at Stanford was supported by the Department of Energy, Office of Basic Energy Sciences, Division of Materials Sciences and Engineering, under Contract No. DE-AC02-76SF00515. F.N. acknowledges support of the European Community through the Marie Curie Fellowship, grant agreement No.~659653. J.F. and E.S. acknowledge support from QMI, NSERC, and CFI.
\end{acknowledgments}

\bibliography{Bibliography}
\end{document}